# Plate Tectonic Consequences of competing models for the origin and history of the Banda Sea subducted oceanic lithosphere

Christian Heine* , Leonardo Quevedo, Hamish McKay, R. Dietmar Müller
EarthByte Group, School of Geosciences, The University of Sydney, NSW, Australia
Email:Christian.Heine@sydney.edu.au

**Abstract**

The Banda Arc, situated west of Irian Jaya and in the easternmost extension of the Sunda subduction zone system, reveals a characteristic bowl-shaped geometry in seismic tomographic images. This indicates that the oceanic lithosphere still remains attached to the surrounding continental margins of northern Australia and the Bird's Head microcontinent. Major controversies exist between authors proposing an allochthonous or autochthonous origin of the Bird's Head block. These two hypotheses have important significant implications for plate kinematic models aiming to reconstruct the tectonic evolution of the region and the late Jurassic seafloor spreading geometry of this now subducted Argo-Tanimbar-Seram (ATS) ocean basin. Wider implications affect the tectonic configuration of the Tethyan-Pacific realm, the distribution of plate boundaries as well as the shape and size of continental blocks which have been rifted off the northeastern Gondwana margin during the Late Jurassic and are now accreted to the SE Asia margin. We apply structural geology restoration techniques, commonly used to reconstruct the kinematic history of deformed rock units on local to sub-regional scales, to unfold the subducted oceanic lithosphere of the Banda Slab using the 3dMove structural restoration software. Our slab unfolding results help to discriminate between the different hypotheses concerning the origin and kinematic history of the Bird's Head tectonic block. According to our preferred model, the block is required to rotate 20-35° clockwise relative to its present-day position and that the initial geometry of the ATS ocean basin was more or less rectangular shaped. We utilize the open-source plate tectonic reconstruction software GPlates to evaluate currently accepted plate kinematic scenarios for the Jurassic and younger seafloor spreading history north of Australia in conjunction with our restored ocean basin geometry. We conclude that the Bird's Head block is autochthonous to western Irian Jaya, with its western margin being a continental transform margin during the rifting and opening of the ATS ocean.

**Keywords**: Bird's Head; Banda Slab; Structural restoration; Plate tectonics; GPlates; 3dMove; Argo-Seram-Tanimbar Ocean basin

# 1 Introduction

The Banda Arc and *Vogelkop* (the Bird's Head) continental block of western Irian Jaya are prominent features in the complex tectonic assemblage of Southeast Asia (Hall, 2011; Hinschberger, 2005; Pubellier et al., 2003; Hall, 2002; Charlton, 2000; Packham, 1996; Bowin et al., 1980). Subducted oceanic lithosphere, imaged by seismic tomography (Spakman and Hall, 2010; Richards et al., 2007; Das, 2004; Hafkenscheid et al., 2001; Milsom, 2001; Gudmundsson and Sambridge, 1998; McCaffrey, 1989), remains attached to the surrounding continental blocks and larger lithospheric plates and provides important clues for reconstructions of the geometry and possible spreading history of vanished ocean basins around Australia's northern margin and the tectonic evolution of the Tethyan-Pacific realm.

The Banda Arc represents the easternmost continuation of the Sunda subduction zone system and is a curvilinear feature bounded by the northern Australian continental margin (Timor to Tanimbar islands) to the south, the Arafura Sea margin/Aru Trough (Kai islands) to the East and the Bird's Head block in the northeast (Fig. 1; Hinschberger et al., 2005, 2001; Honthaas et al., 1998; Norvick, 1979). Marine magnetic surveys and dredging in the Banda Sea confirmed the Neogene age and back arc origin of the oceanic crust in the Wetar and Damar Basins (Hinschberger et al., 2005; Honthaas et al., 1998).

The Wadati-Benioff zones of several regional subduction zones and the overall geometry of slabs in the mantle beneath this region have been imaged by several local and regional seismic tomographic studies over the past decade (Spakman and Hall, 2010; Richards et al., 2007; Das, 2004; Hafkenscheid et al., 2001; Milsom, 2001; Gudmundsson and Sambridge, 1998; McCaffrey, 1989). These studies indicate that the Banda Slab is bowl-shaped, with its deepest parts bottoming out on the 660 km transition zone below southeastern Sulawesi (Spakman and Hall, 2010; Richards et al., 2007; Das, 2004; Gudmundsson and Sambridge, 1998; McCaffrey, 1989; Fig. 2). They also suggest that the slab is attached to the lithosphere of the surface plates surrounding the Banda Arc, namely the Australian plate in the South and East and the Bird's Head tectonic block in the Northeast (Spakman and Hall, 2010; Fichtner et al., 2010; Richards et al., 2007; Hinschberger et al., 2005). The western boundary of the northern limb of the Banda slab has a relatively sharp boundary,

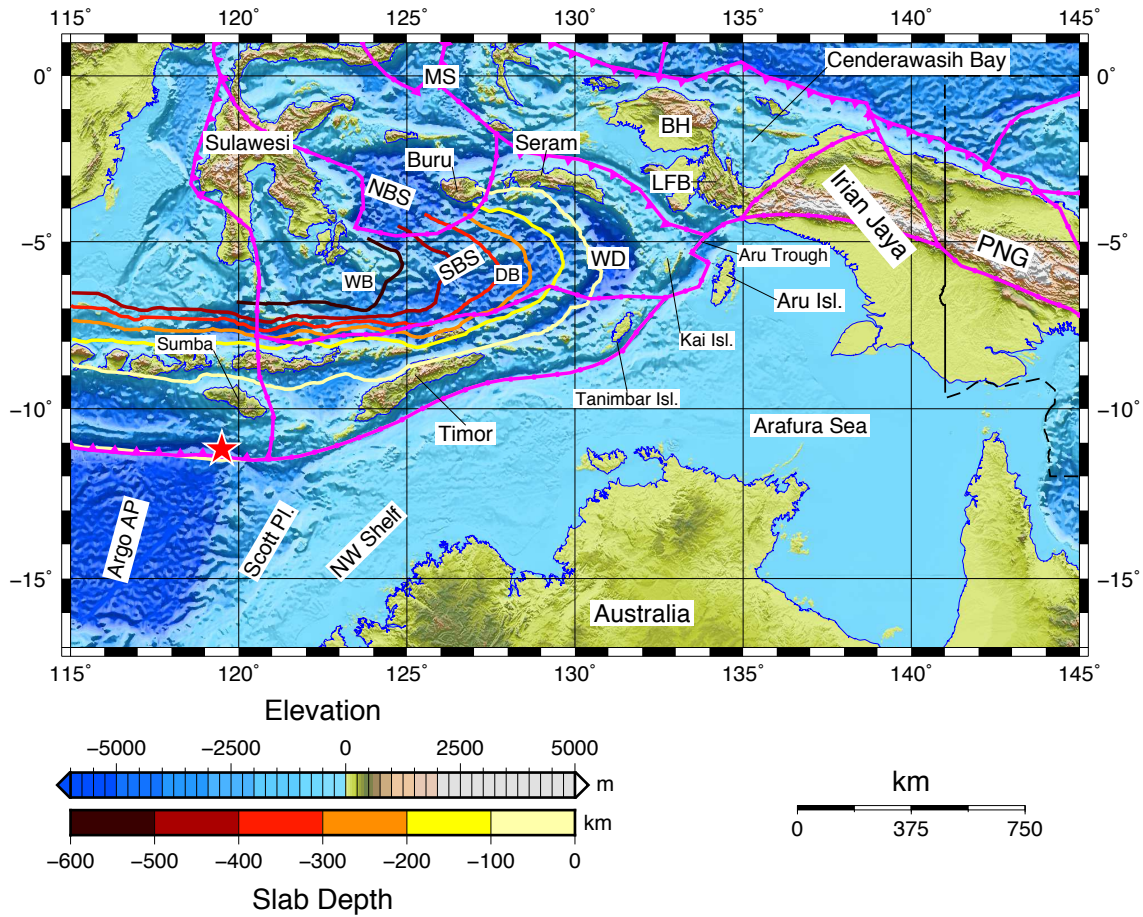

**Figure 1:** Regional overview map. Show is bathymetry/elevation from ETOPO1 data with plate boundaries (Bird, 2003) as magenta-colored lines. Contours of Benioff zone at 100 km depth intervals from Spakman and Hall (2010) are shown as solid color-coded lines. Red star indicates location of our preferred unfolding pin surface centerpoint. Abbreviations: BH: Bird's Head; DB: Damar Basin ; LFB: Lengguru Fold Belt; MS: Molucca Sea; NBS: North Banda Sea; SBS: South Banda Sea; PNG: Papua New Guinea; WB: Wetar basin; WD: Weber Deep

described as a "subvertical slab tear" separating it from the Molucca Sea slab beneath the Molucca Sea (Fig. 1) to the north (Richards et al., 2007). The oceanic lithosphere now subducted beneath the Banda Arc and eastern Sunda Subduction zone system (encompassing the Sunda and Banda subduction zones) is part of the Indo-Australian plate and was generated through oceanic seafloor spreading related to the last rifting event along the northern Gondwana margin in the late Jurassic (Gibbons et al., 2012; Heine and Müller, 2005; Heine et al., 2004; Fullerton et al., 1989). Its seafloor spreading history can be reconstructed through the magnetic reversal patterns preserved in the Argo Abyssal Plain off Northwest Australia (Gibbons et al., 2012; Heine and Müller, 2005; Heine et al., 2004; Fullerton et al., 1989; Heirtzler et al., 1978). This piece of oceanic crust is one of the oldest in the Indian Ocean and, together with observations from basins subsidence on the Northern and Northwestern Australian margin, documents a rifting event in the latest Jurassic and the subsequent removal of a continental sliver (e.g. Heine and Müller, 2005; Longley et al., 2003; von Rad et al., 1989).

The Bird's Head, the westernmost part of Irian Jaya, is a stable continental block surrounded by highly active plate boundaries (Bock et al., 2003; Bird, 2003; Stevens et al., 2002; Kreemer et al., 2000; DeMets et al., 1994; Ekström and England, 1989; Dow and Sukamto, 1984; Mc-Caffrey, 1989) and has Gondwana-type basement overlain by Paleozoic to Mesozoic sediments, akin to those found on the northern Australian margin (Decker et al., 2009; Pubellier et al., 2003; Baillie et al., 2004; Charlton, 2000; Packham, 1996; Pigram and Panggabean, 1984). Three major tectonic boundaries delimit the block, i.e. the Sorong/Yapen Fault zone in the north, the Tarera Aiduna Fault Zone in the SE and the Seram trench as part of the Banda Subduction zone (e.g. Bock et al., 2003).

The affinity of the older sedimentary sequences and the basement of the Bird's Head block with the northern Australian margin has led to controversies about the origin of the Bird's Head. Many authors propose that the Bird's Head is allochthonous and has rifted off the eastern or northern Australian margin in the Jurassic (Pubellier et al., 2003; Charlton, 2001; Packham, 1996; Giddings et al., 1993; Pigram and Symonds, 1991; Pigram and Panggabean, 1984) whilst others favour no major relative motion between the Bird's Head and western Irian Jaya (Hill, 2005; Baillie et al., 2004; Hill and Hall, 2003; Metcalfe, 1996). Published tectonic reconstructions for the late Jurassic of the region are unclear on how the "Proto-Banda" Ocean (Hall, 2002) was formed. However, the formation and initial geometry of the Banda Embayment or the Argo-Tanimbar-Seram ocean provides key boundary conditions on the origin of the Bird's Head block as well as on the post-Jurassic seafloor spreading history north of NE Gondwana and plate kinematic history of terranes derived from this margin (e.g. Gibbons et al., 2012; Hall, 2011; Heine et al., 2004).

Charlton (2001) proposes an Early Permian to Middle Triassic extension and rifting event (70-25 Myr duration, depending on the stages chosen), which translated the "Greater Sula Spur" continental block, including the Bird's Head, from a position NE of Timor to its current position, opening his "Meso-Tethys 2" ocean. This rifting event left exhumed basement as "Banda Embayment Terrane" which was unmodified in subsequent rifting events (Charlton, 2001). Recent work by Hall (2011) has the "Banda Blocks" rifting from the Banda Embayment in the late Jurassic, contemporaneously with the rifting and seafloor spreading in the Argo Abysssal Plain. In his reconstructions the Bird's Head is part of the "Greater Sula Spur", a narrow continental promontory, which is autochthonous to Irian Jaya. The Bird's Head is fixed relative to Australia and western Papua in present day coordinates and the seafloor spreading north of Australia is modeled first as single spreading system with NW-SE opening and from 135 Ma onwards as triple junction system between India, Australia and the Incertus/Woyla Arc/SE Asian margin. The plate motion history implied by his reconstructions to open the Proto-Banda Sea, however, remains to be tested. Our slab restoration analysis presented in this paper aims to test the different hypotheses concerning the origin of the Bird's Head and aims to provide a robust template for restoring the geometry of the former Argo-Tanimbar-Seram ocean basin.

## 2 Slab unfolding

We assume that the Benioff Zone contours around the Banda slab (Spakman and Hall, 2010) represent the surface of the subducted oceanic lithosphere as an approximation and that no major internal deformation has affected the geometry of the slab. However, Das (2004) suggested that parts of the Banda Slab show localised slab thickening, in particular in the region of highest slab curvature beneath the eastern part of the Banda Arc and subsequently proposed slab folding or a second slab as the cause. Other authors (Spakman and Hall, 2010; Richards et al., 2007; Gudmundsson and Sambridge, 1998; McCaffrey, 1989; Cardwell and Isacks, 1978) argue against 2 different slabs and show the Banda Slab as continuous feature (Fig. 1 and 2). Earlier models agree well with the contours of the Wadati Benioff Zone by Spakman and Hall (2010). The Slab1.0 model (Hayes et al., 2012) does not include the Banda Arc. For the purpose of this paper we use only the slab surface in our restoration models.

### 2.1 Methodology

The digitised Benioff zone contours were converted into a triangulated surface mesh using Midland Valley's 2D and 3D Move structural restoration software, hereafter termed 'slab surface'. We scaled the absolute coordinates of the features by 1000 as the software has limited abilities to deal with depths in the 100 km range. Once the slab surface had been generated, it was then unfolded to a target surface, here a plane horizontal surface at 10 km depth, using a flexural slip algorithm (Griffiths et al., 2002; Gratier et al., 1991; Rouby et al., 2000; Gratier and Guillier, 1993). We tested different unfolding pin surface orientations and unfolding directions (cf. Table 1). The unfolding planes represent the direction of strain when restoring a folded surface. We stress that this method cannot account for internal deformation of the slab, and in addition we only use the slab surface, not a volumetric definition (ie top/base) of the slab and the unfolding is carried out in a rectangular UTM Zone 52 M projection. For the case of the Banda Arc, this approach can be justified as the lateral extent of the Arc system is about 800 km in North-South direction and about 1500 km in East-West direction.

Our preferred anchor point for the restoration is located at the northeasternmost corner of the Argo Abyssal plain, at the junction between the Sundaland trench, the northern prolongation of the Scott Plateau and the Indian Ocean lithosphere (red star in Fig 1). The location is the easternmost surface occurrence of the subducting oceanic slab of the Indo-Australian plate. We also evaluated other unfolding pin surface locations on the trench between Timor and Tanimbar island (Models 1-3) as well as around the Kai/Aru islands in the apex of the Banda Arc (Model 7a). The unfolding pin location for models 4--9 is shown as red star in Fig.1.

**2.2 Restoration models**

We have tested a number of different parameters to restore the geometry of the Banda slab to 10 km depth using flexural slip unfolding, mainly by varying the unfolding plane direction and the pin surface location. In all cases, our unfolding plane is oriented orthogonally to our pin surface. As the original slab surface is non-cylindrical, the flexural slip unfolding mechanism will not yield unique results depending on the choice of these parameters. We used 3dMove's "Cylindrical Analysis" Tool to compute the orientation parameters of our digitised slab surface. The average cylindrical vector of the slab surface has an azimuth of $276°$ and plunge of $11°$.

The geometry of ocean basins and their adjacent margins formed by continental rifting and subsequent breakup is largely dependent on the orientation of the minimum horizontal stress at the time of opening and initial spreading. Most present-day oceanic basins generated by continental breakup are elongated with a spreading direction oriented roughly perpendicular to the present-day continental margins. Exceptions are highly oblique or transfer margins such as in the Equatorial Atlantic domain, the Gulf of California or the Gulf of Aden.

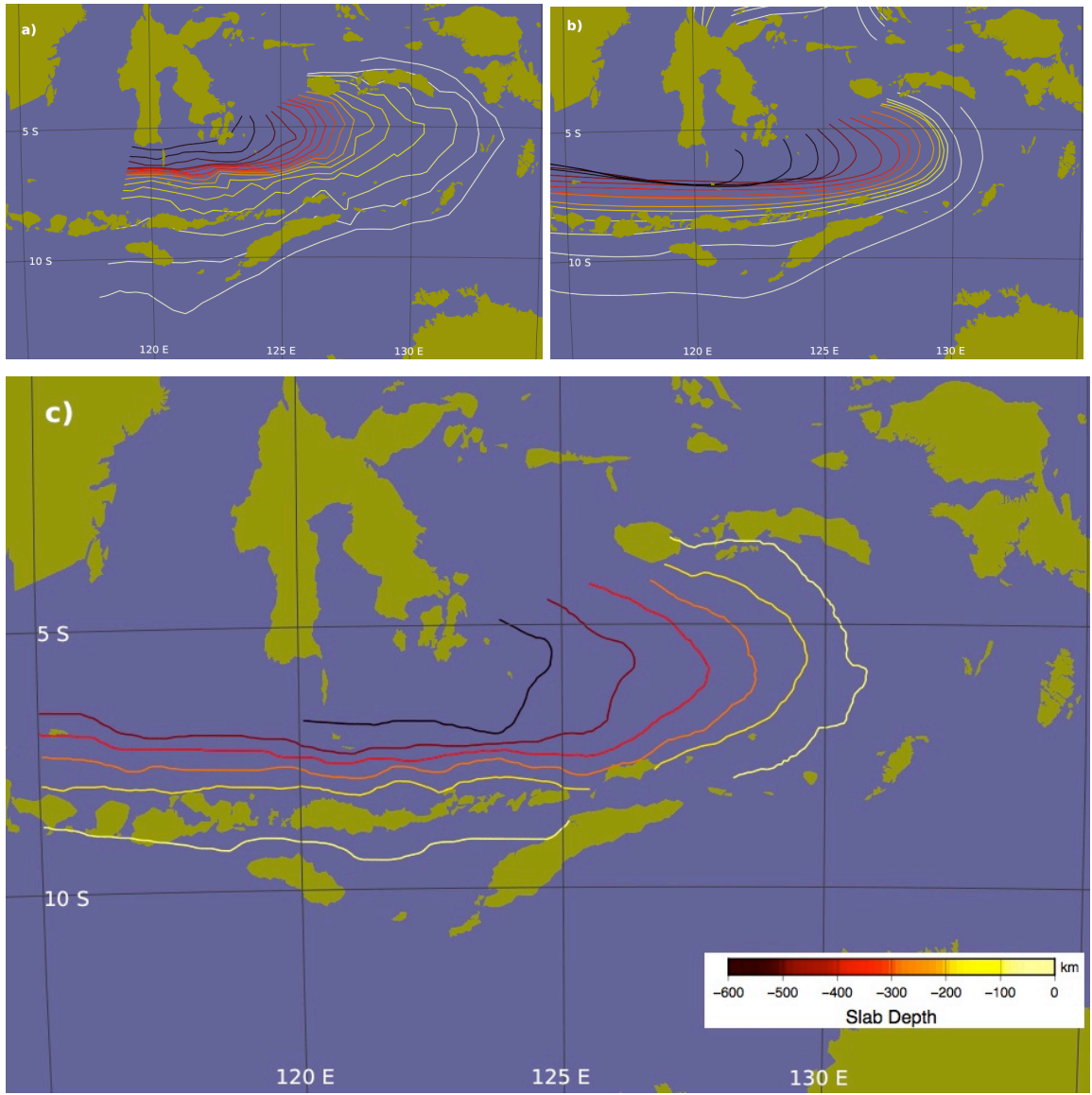

**Figure 2:** Slab models for the eastern part of the Sunda Slab as contour lines visualised in GPlates with land areas (green) and ocean (blue): a) Richards et al. (2007), b) RUM model of Gudmundsson and Sambridge (1998), c) Spakman and Hall (2010). Colorcode for slab contours is the same as in Fig. 1. We use the term "Banda Slab" to describe the easternmost, highly curved part of the Sunda slab which extends from the Burma region towards eastern Indonesia. In general, the models show good agreement in both geometry and depth to the Benioff Zone (here used as indicator for the top of the slab).

Our hypothesis is that breakup and seafloor spreading which generated the now subducted oceanic basin north of Australia, here called the Argo-Tanimbar-Seram (ATS) ocean basin, should have occurred similarly to existing oceanic basins, i.e. roughly perpendicular to the continental margins bounding the ocean basin at present day. Parts of the continental margin of that ATS ocean are preserved in-situ as the continent-ocean boundary along the Argo Abyssal Plain off northwest Australia. This constrains the possible spreading geometries between the currently accepted opening direction of the Argo Abyssal Plain (D1) and the strike of its eastern margin (D3).

The first unfolding direction (D1) is based on currently modelled initial spreading directions in the Argo Abyssal Plain. The Argo Abyssal Plain is one of the last pieces of in-situ oceanic crust which recorded the late Jurassic rifting and breakup history along the northern Australian margin (Gibbons et al., 2012; Heine and Müller, 2005; Heine et al., 2004; Fullerton et al., 1989). The breakup and subsequent seafloor spreading occurred at an orientation of roughly N30°W.

The most extreme end-member direction (D4) is assumed to be parallel to the strike of the eastern margin of the Aru Trough. Such direction would be required if the Bird's head Block had to originate from the Timor to Tanimbar segment of the northern Australian margin (Charlton, 2001). The Aru Trough represents a tectonically young extensional graben west of the Aru Islands in the easternmost part of the Banda Arc (Jongsma et al., 1989; Charlton et al., 1991; Untung, 1985). We hypothesise that the Aru Trough and its northern prolongation could have marked a possible transform margin for the ATS ocean. In such a scenario, the rifting along the northern Australian margin would most likely have occurred in two different compartments, separated by a triple junction north of the Scott Plateau. Such a configuration would have been active starting in the Late Jurassic, contemporaneously with or slightly after the onset of rifting or seafloor spreading in the Argo Abyssal Plain. While marine magnetic anomalies confirm an opening of the Argo Abyssal Plain in NW-SE'ly direction (e.g. Gibbons et al., 2012; Heine & Müller, 2005; Mihut & Müller, 1998; and references therein), the eastern branch of the triple junction spreading ridge could have developed north of the Scott plateau, trending at about 120° (or N150E) to the Argo spreading direction and translating the Bird's Head block along the Aru Trough margin to the NE to its present position. In the reconstructions by Hall (2011) seafloor spreading occurs sub-parallel to the Argo spreading (NW-SE extension) and rifts a "Banda Block" out of the Banda embayment along a Bird's Head transform margin Late Jurassic times (150 Ma). While his reconstructions are very similar to the quantified plate kinematic models by Gibbons et al., (2012) or Heine & Müller (2005), the reconstructions shown in Hall (2011) for 150 Ma and 130 Ma imply significant compression between the Argo and Banda Blocks between 150 and 130 Ma. Additionally, the shape of his "Greater Sula Spur" will have caused significant compression along the Bird's Head transform margin. In our hypothetical triple junction scenario, the triple junction would have evolved at the same time or slightly after the rifting in the Argo Abyssal Plain in the Late Jurassic. The unfolding direction chosen for such scenario is N40°E. Intermediate unfolding directions are subparallel to the Scott Plateau margin of the Argo Abyssal Plain and trend N15°E (D2) to N25°E (D3) as well as orthogonal to the computed average strike of the subducting Banda Slab (N84°E).

Additional intermediate cases are listed in Table 1. Variations in the location of the pin surface as well as a 10° dip of the unfolding plane have been tested in accordance with the results of our orientation analysis. For all our models we assume that the Bird's Head tectonic block is connected to the subducted oceanic slab around the Seram segment of the Banda Arc and that there was not differential motion or tearing between subducted lithosphere and surface platelet.

| Model No. | Pin surface orientation | | | | Unfolding plane |
|---|---|---|---|---|---|
| | x | y | z/datum | strike/dip | strike/dip |
| 4 | -465.862 | 8715.426 | -10/-10 | N15°E/0 | N105°E/0 |
| 5 | -465.862 | 8715.426 | -10/-10 | N25°E/0 | N115°E/0 |
| 6 | -465.862 | 8715.426 | -10/-10 | N40°E/0 | N130°E/0 |
| 7 | -465.862 | 8715.426 | -10/-10 | N240°W/0 | N330°E/0 |
| 7a | 50.000 | 8850.000 | -10/-10 | N240°W/0 | N330°E/0 |
| 8 | -465.862 | 8715.426 | -10/-10 | N90°E/0 | N0°E/0 |
| 9 | -465.862 | 8715.426 | -10/-10 | N186.43°E/0 | N276.43°E/10 |
| 10 | 0 | 9400.000 | -10/-10 | N186.43°E/0 | N276.43°E/10 |

**Table 1:** Different flexural unfoding models used to flatten and restore the Banda Slab. Easternmost extend of oceanic crust of the oceanic crust: x = -465.862, y = 8715.462; z = -10. Model 7 like Model 6 but with algorithm tuned for steep beds.

## 3 Results

Based on the workflow described in the previous section we have constructed a set of restored Banda Arc slab geometries (Figs. 4 & 5). Our attempts to restore and unfold the subducted part of the slab using different unfolding pin surface locations and unfolding directions clearly indicate that the choice of these parameters has a key impact on the shape of the unfolded slab, given the constraints on how the flexural slip algorithm is implemented in the 3dMove software. The flexural slip restoration algorithm attempts to preserve length and area of the unfolded surface. This assumes that there is no internal deformation which has affected the slab surface, an assumption which is most likely not correct for the tightly folded inner part of the Banda Sea slab. Figure 3 shows that the change in area between the original slab surface in our digital model ("Original") and different restored slab geometries (Models 4–9) in 3dMove are indeed only ranging between 0.8 and 3.69 %. Here, model 4 represents the least change in area between the original slab and the restored surface (+0.81%), and Model 6 the largest increase in surface area (+3.69%) compared to the original slab surface. We use the area mismatch to discriminate between the best fit models and the most likely restored slab geometry. Models 4, 5, and 9 result in the least amount of area change between original and restored slab surface.

The restored geometries of the slab surface using different choices of unfolding parameters are shown in Fig. 4. The unfolding direction for the best-fit models varies between N6.43°E (Model 9) and N25°E. The restored slab geometries clearly indicate that the Bird's Head/Seram slab segment is required to rotate significantly in a clockwise fashion to account for the observed present-day slab geometry (Fig. 5). Additionally and perhaps most interestingly, significant lateral displacement is implicit from our results, requiring the Bird's Head/Seram slab segment to move northward, going back in time. We have computed the displacement for the easternmost 70 vertices of the 100 km depth slab contour and the end vertices of each contour in all models. A best fitting great circle and average position for the vertices was computed by minimising the sum of squares of cosines of angular distances of the individual vertex points. The results are an abstraction of the earlier restoration exercise. The results for our preferred models (4, 5 and 9) indicate that the Bird's Head microcontinent/Seram slab segment needs to rotate by between 20° (Model 9), 28° (Model 4) and 35° (Model 5) degrees clockwise, going backward in time. The amount of lateral displacement ranges between 177 km for Model 9 (Fig 5f) and 458 km in Model 6 (Fig. 5c). Our best-fit models require a NNE to northerly displacement of 177 km (Model 9), 300 km (Model 4), and 370 km (Model 5) when comparing the length of the present-day Seram slab segment and the restored Seram slab segment (Fig. 5). All our results indicate that the geometry of the embayment of the ATS ocean was most likely rectangular shaped, with the Seram segment of the Banda Arc oriented semi-orthogonal

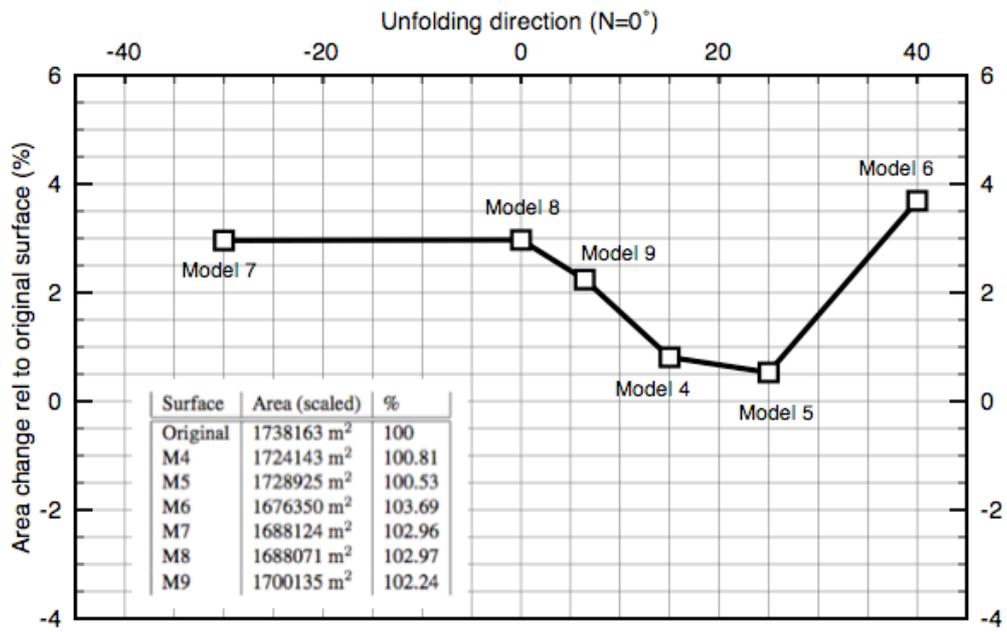

**Figure 3:** Plot and tabulated data of area change versus unfolding plane direction. Area change is given in % relative to original slab surface. Unfolding plane orientation is given in degrees relative to North. Models 5, 4, and 9 show the smallest increase in area of the restored slab relative to the original surface.

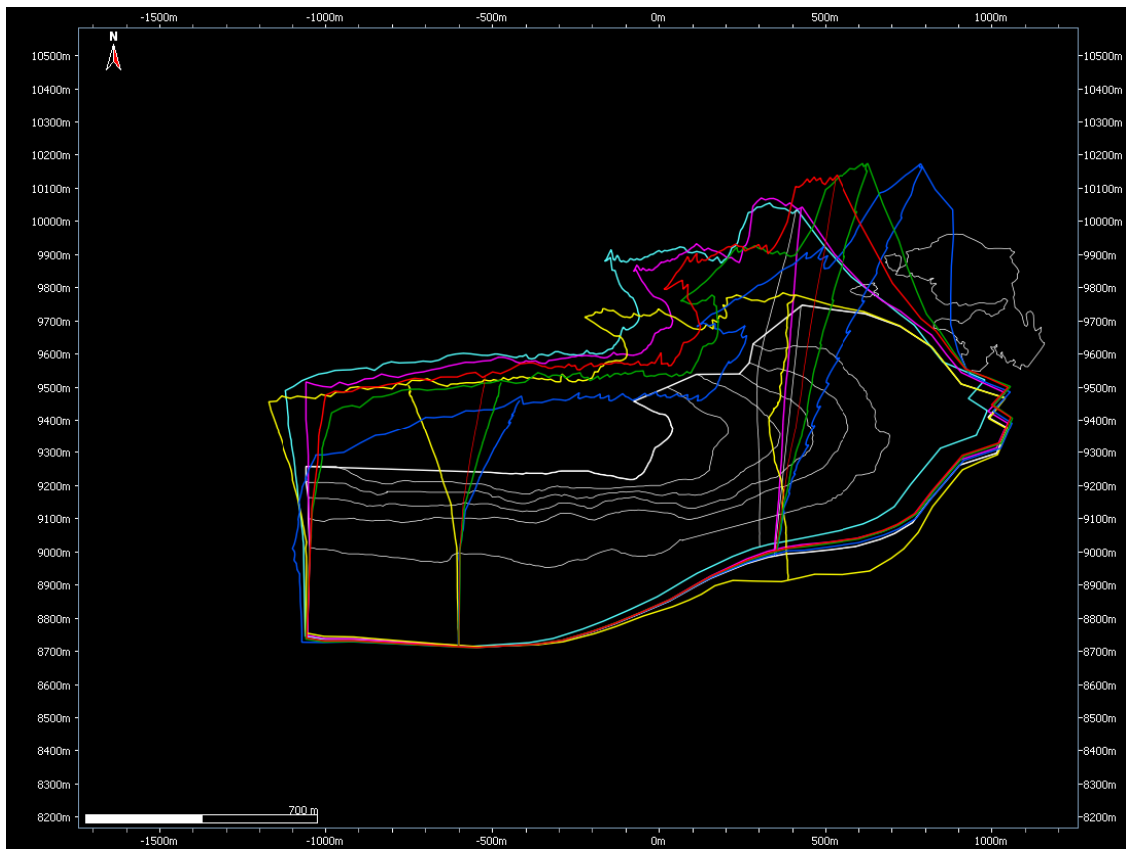

**Figure 4:** Map view of the results of slab unfolding using flexural slip algorithm for different restoration scenarios 4-9 as indicated in Table 1. White outline represent original Benioff zone depth contours in 100 km intervals as well as the Bird's Head present day coastline of the Bird's Head, Misool and Salawati Island based on the GSHHS database (Wessel and Smith, 1996). Restored Benioff zone contours are colorored according to restoration parameter setup: Model 4 -red; Model 5 -green; Model 6 -blue; Model 7: yellow; Model 7a: apricot; Model 8: magenta; Model 9: cyan. Apart from unfolding scenario of Model 7 (yellow), all models require lateral displacement and/or rotation of the Bird's Head relative to the Australian northern margin/Tanimbar-Kai segment. This is either due to the Seram slab segment rotating clockwise (Models 4, 5, 6, 8 & 9) or the Tanimbar-Kai slab segment being displaced southward (Model 7) in the different restoration scenarios. Thin straight white line is a section marker on the 3D slab surface. Coordinate system is UTM 52M scaled by 1000.

# 4 Discussion

Our slab restoration and unfolding models indicate that the subducted oceanic lithosphere of the Argo-Tanimbar-Seram Ocean which now forms the Banda Slab can be restored using structural geology techniques. By testing a range of unfolding scenarios using flexural slip restoration algorithms in 3D, we are able to arrive at a set of reasonable restored slab geometries that could have represented the pre-subduction geometry of the Jurassic ocean basin north of Australia (Figs. 4 & 5). A statistical evaluation of the area changes between the present-day modelled slab surface area and the restored surface area indicates that our models 4, 5 and 9 are the most likely reconstructed geometries of the unfolded slab (Fig. 3). The minimum area change is achieved with an unfolding direction oriented N25°E and the unfolding pin surface oriented orthogonally to it. The flexural slip restoration will cause a clockwise rotation of the Bird's Head block by about 28° (Model 4)–35° (Model 5) going back in time (Fig. 5). The amount of rotation from model 5 is in accordance with the 30° to 40° of rotation proposed by Charlton (2000). The final restored position resembles the total integrated kinematic history which allows to restore a version of the initial shape of the ATS ocean based we believe is robust (Figs. 5 & 6).

We have used a recent tomographic study (Fichtner et al., 2010) to model the shape of the extended continental lithosphere around Australia from the Scott Plateau to Tanimbar (Thick red line in Fig. 6). In combination with our Banda slab restoration results we are able to reconstruct the geometry of the Jurassic ocean basin and the continental margins bordering the ATS ocean (Fig. 6). Late Jurassic extension and subsequent seafloor spreading along the North West Shelf was predominantly oriented in a NW-SE direction (Longley et al., 2003), confirmed by interpretations of the marine magnetic anomalies in the Argo Abyssal Plain (Gibbons et al., 2012; Heine and Müller, 2005; Heine et al., 2004; Fullerton et al., 1989). If we assume a lateral continuation of the orientation of the Late Jurassic stress field, then the restored continental margin of the Australian plate from the

Scott Plateau to the apex of the present-day Banda Arc around the Aru Islands is a normal passive margin. The Bird's Head/Seram segment of the Banda Slab becomes a continental transfer margin, akin to the Keta-Togo-Benin margin segment in the Equatorial Atlantic. Seismicity in the western Arafura Sea indicates thick continental lithosphere and rare occurrences of upper continental lithospheric mantle earthquakes, which point to a cratonic origin underlying the region east of the Aru islands and the Arafura Sea (Sloan and Jackson, 2012). It is likely that western parts of Irian Jaya and the probably event the Bird's Head are also underlain by similar lithosphere, which could explain the tectonic rigidity of this block is such a highly dynamic tectonic setting.

Using currently accepted spreading scenarios (Gibbons et al., 2012; Heine and Müller, 2005; Heine et al., 2004; Fullerton et al., 1989) for the ATS ocean basin formation, geometrical constraints do not allow the Bird's Head tectonic block to be derived from the northern Australian margin during the late Jurassic rifting. Instead the block needs to be autochthonous to western Irian Jaya. The rifting event, which affected the whole northern Australian margin from the Exmouth Plateau to the Arafura Sea has removed continental blocks from the northern Australian margin, which are now accreted and incorporated into the SE Asia (Gibbons et al., 2012; Metcalfe, 2011; Heine and Müller, 2005; Metcalfe, 1991, 1994). The regional stress field associated with the rifting of such a block, most likely Hall's Southwest Borneo block (Hall, 2011), will also have affected the Bird's Head. Hence we propose that the Bird's Head occupied an analogous position as the Exmouth Plateau on the Northwest Australian margin, which is an area that experienced widespread extension during the late Jurassic and early Cretaceous rifting along the northern and western Australian margins, respectively (Longley et al., 2003; von Rad et al., 1989). The thermal subsidence related to the break up event would have occurred slightly later than on the northern Australian margin due to the lateral offset along the Bird's Head/Seram transfer margin.

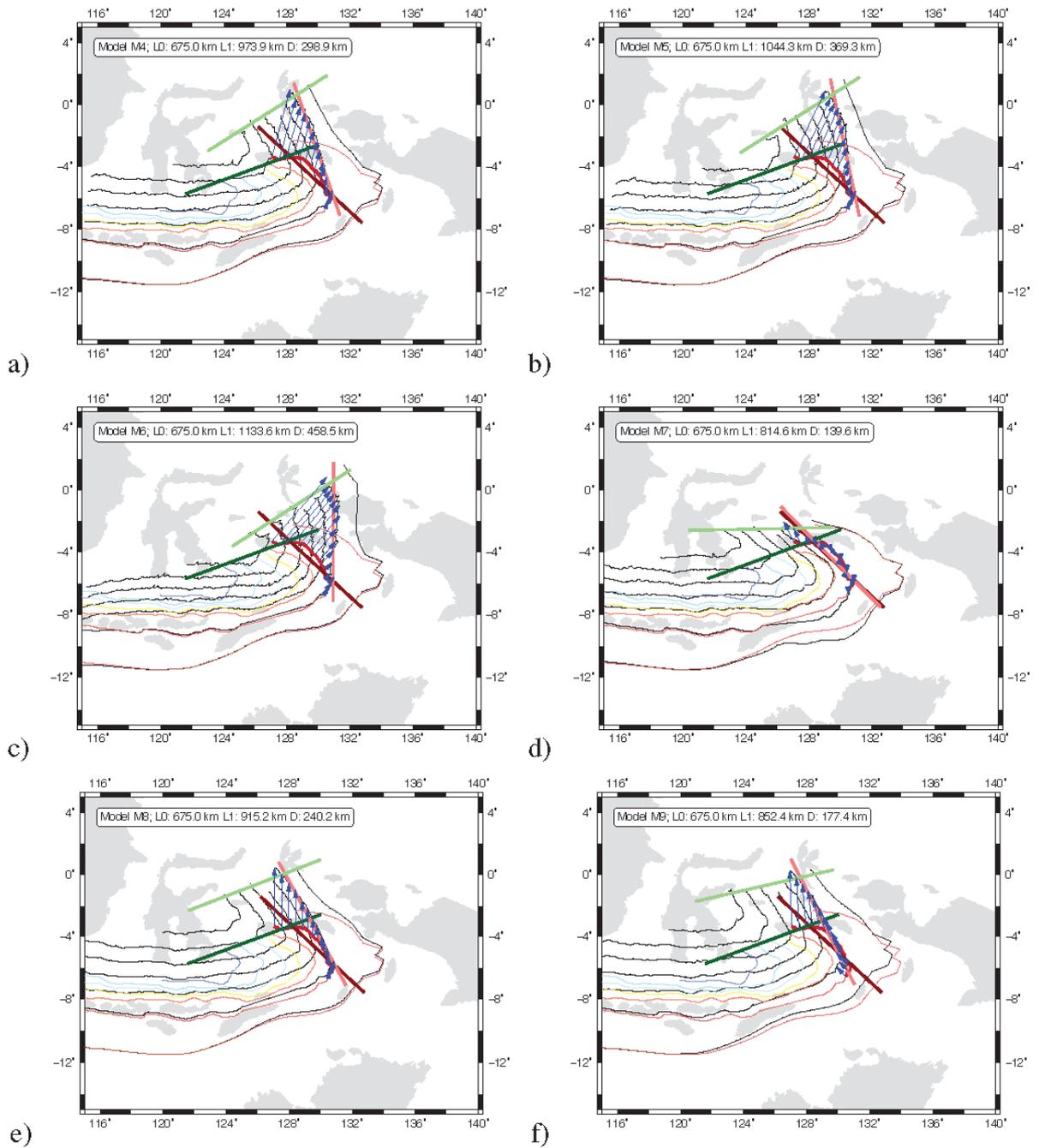

**Figure 5:** Restored positions of the 100 km slab contour in different restoration scenarios using a flexural slip algorithm with different unfolding directions (see Table 1 for different model parameters). Present-day slab contours based on Spakman and Hall (2010) are coloured in 100 km depth contours. Restored slab contours are plotted as solid black lines. Text inset gives original length of a 70-vertex long line segment representing the Bird's Head slab segment (L0), the restored segment length (L1) and amount of extension (D). Vectors indicate the displacement direction between original and restored line segments. Dark red and dark green are the location of the fitted great circle segment to the 70 vertex-long Seram slab segment and the end vertices of the individual depth contours, respectively. Light red and light green lines indicate the position of those two features using the different restoration parameters. Maps are in Mercator projection.

Our restoration model only allows us to reconstruct the initial geometry of the late Jurassic ATS ocean basin based on the currently observed slab contours. The detailed kinematic history of the Bird's Head microplate is clearly governed by two main motions since the northern margin of the Australian plate reached the Sunda subduction zone. Firstly, it is the rotation of the Bird's Head due to the slab rolling back and sinking, and secondly a southward translation due to the continued northward motion of Australia, leading to compression, concentrated in the apex of the Banda Arc Slab. The latter could be responsible for the enigmatic seismicity pattern in the easternmost Banda Arc, which led Das (2004) to conclude that the Banda Slab is composed of two individual pieces of subducted oceanic lithosphere. Based on existing plate reconstructions and synthesised regional geological data, the rollback of the Banda Slab and the formation of the currently observed slab surface is a complex and dynamic process which is governed by the relative motions of the larger plates in this area (Spakman and Hall, 2010; Hill, 2005; Hill and Hall, 2003; Hall, 1996).

## 5 Conclusions

We have demonstrated that the restoration of the Banda Slab in the eastern Indonesia subduction zone system using flexural slip restoration methods has important consequences for restoring the geometry of the margins of the late Jurassic Argo-Seram-Tanimbar ocean basin and its seafloor spreading history. Our reconstructed slab geometry requires that the Bird's Head block is autochthonous to western Irian Jaya and can certainly not be derived from the eastern Australian margin. An origin of the Bird's Head block from the Timor-Tanibar section along the northern Austalian margin is also very unlikely as it requires a complex triple-junction scenario, translating the Bird's Head northeastward along the Aru trough. Based on currently accepted plate kinematic models using marine magnetic anomalies in the Argo Abyssal Plain and the rules of plate tectonics as proxy for the opening ATS ocean basin, the "north Australian margin origin" hypothesis is very unlikely, but cannot be ruled out with certainty. Our modeling further indicates that a model consisting of two independent slabs for the subducted oceanic crust in the beneath the Banda Arc, as proposed by Das (2004) is geometrically and tectonically unfeasible. Our favoured restoration model requires a $35°$ clockwise rotation of the Bird's Head and slight northward displacement relative to its present-day position.

Lithosphere thickness, basement character and overlying Paleozoic to Mesozoic sediments clearly indicate a similar Post-Triassic tectonic history of the Bird's Head compared to the northern Australian margin (Decker et al., 2009; Baillie et al., 2004; Packham, 1996). Our reconstructed continental margin geometry around the ATS ocean and paleo-stress field orientations from plate kinematic models indicates that the Timor to Tanimbar/Kai Islands segment of the northern Australian margin was likely a passive continental margin. The Bird's Head segment of the Banda Arc, from the Kai Islands to Seram is a transform margin in this tectonic scenario. We consider the tectonic position of the Bird's Head to be analogous to that of the Exmouth Plateau of the Australian North West Shelf.

Since the collision of the leading edge of the Australian continental lithosphere with the Sunda Trench at about 15 Ma, the slab rollback into the easternmost "Gulf of Argo-Tanimbar-Seram" has caused a complex kinematic history of the Bird's Head. The block rotated counterclockwise, opening the Cenderawasih Bay, while being slightly translated southward, creating the Lengguru Fold Belt. Our slab restoration only yields the initial geometry and integrated kinematic history of the ocean basins and tectonic elements attached to the oceanic lithosphere of the Banda Slab. Future work will include a quantified, detailed late Mesozoic to Cenozoic plate kinematic reconstruction of this area.

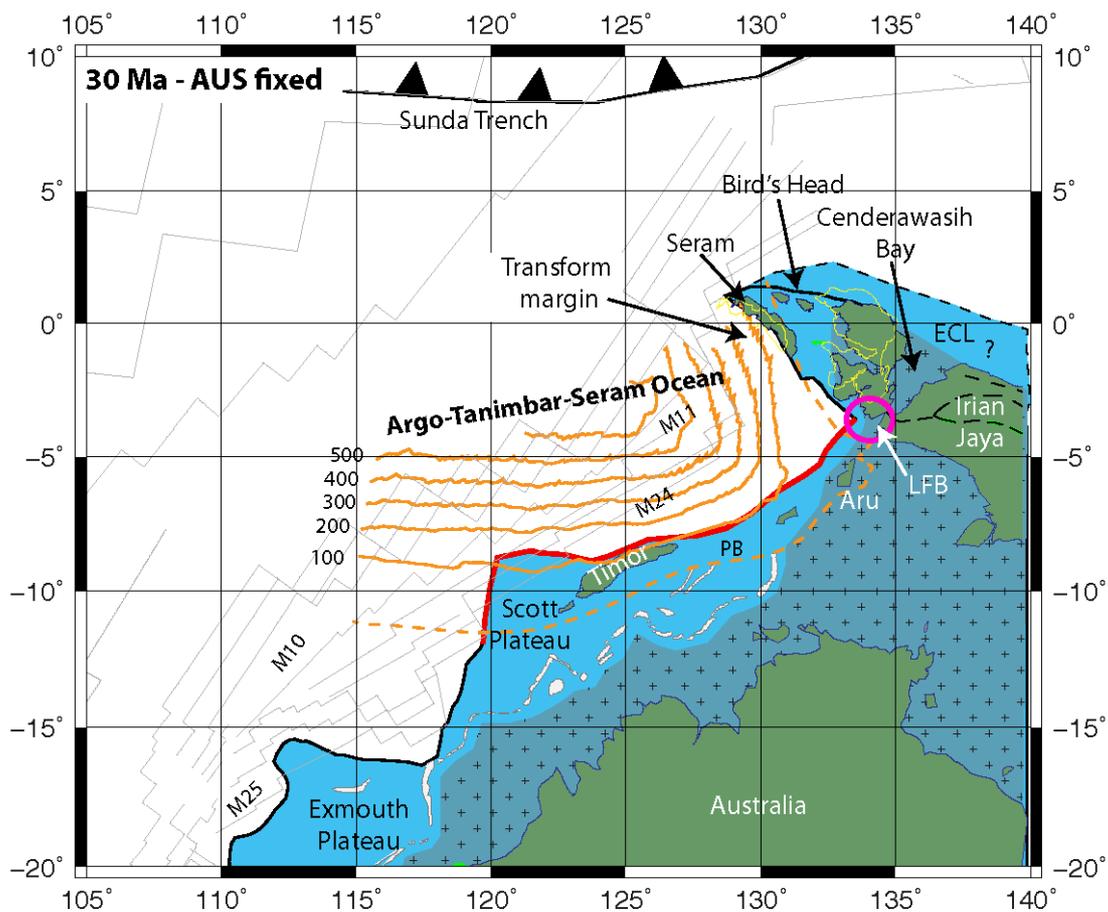

**Figure 6:** Reconstruction of the Argo-Tanimbar-Seram Ocean basin at 30 Ma, with Australia fixed in present-day coordinates using restoration model 5. Gray-blue colors with cross patterns indicate unthinned continental lithosphere of the Australian Plate. Light blue shading represents likely distribution of extended continental lithosphere from Late Jurassic Rifting event. Thin yellow outlines

indicate present-day coastline position of the Bird's Head and Seram using restoration Model 4. Numbers indicate present-day depth contours for restored slab geometry Model 5. Interpreted isochrones based on marine magnetic anomalies M25 (154.1 Ma), M24 (152.1 Ma), M11 (133.3 Ma), M10 (130.9 Ma) are based on Heine et al. (2004). Thick red line indicates restored extend of Australian continental lithosphere based on Fichtner et al. (2010). LFB: Future position of the Lengguru fold belt indicated by circle; PB: present-day plate boundary. Light blue indicates extended continental lithosphere, gray-blue colours with cross signature indicates undeformed continental lithosphere, green indicates present-day emergent land. Light grey polygons on NW Shelf are stranded Jurassic basins (M. Norvick, pers. comm and Heine and Müller (2005)) ECL: Extended continental crust north of Papua New Guinea (Mamberano Basin after Pubellier et al., 2003; and J. Suppe, pers. comm. 2012). For the restoration we have assumed no internal deformation of the Bird's Head.

## Acknowledgements

Midland Valley is acknowledged for an acedemic software license to the University of Sydney. Figures were generated using the Generic Mapping Tools (http://gmt.soest.hawaii.edu) and GPlates (http://www.gplates.org). Christian Heine and Leonardo Quevedo are funded by ARC Linkage Project L1759 with Shell E&P, and TOTAL. We acknowledge John Suppe and Jonny Wu (Taiwan National University) and Sabin Zahirovic (USYD) for constructive comments during the first TaiSy workshop. Jasper Zoethout (Statoil) is thanked for an initial restoration and 3DMove setup and discussions. Eric Blanc and Øyvind Engen (Statoil) are acknowledged for feedback in the early phase of this project.